\documentclass[11pt,aps,nofootinbib,prd,aps,epsf,floats,axodraw,amsmath,amssymb,amsfonts]{revtex4}
\usepackage{amsmath, amssymb}
\newcommand{\mathsym}[1]{{}}

\usepackage{graphicx}% Include figure files
\usepackage{amsmath}
\usepackage{amssymb}
\usepackage{bm}% bold math
\newcommand{\be}{\begin{equation}}
\newcommand{\ee}{\end{equation}}
\newcommand{\bea}{\begin{eqnarray}}
\newcommand{\eea}{\end{eqnarray}}

\newcommand{\rem}[1]{}
\newsavebox{\PSLASH}
 \sbox{\PSLASH}{$p$\hspace{-1.8mm}/}
 
\renewcommand{\theequation}{\thesection.\arabic{equation}}
\newcounter{saveeqn}
\newcommand{\add}{\addtocounter{equation}{1}}
\newcommand{\alpheqn}{\setcounter{saveeqn}{\value{equation}}%
\setcounter{equation}{0}%
\renewcommand{\theequation}{\mbox{\thesection.\arabic{saveeqn}{\alph{equation}}}}}
\newcommand{\reseteqn}{\setcounter{equation}{\value{saveeqn}}%
\renewcommand{\theequation}{\thesection.\arabic{equation}}}

 \newsavebox{\notrightarrow}
 \sbox{\notrightarrow}{$\to$\hspace{-4mm}/}
 
 \newsavebox{\PARTIALSLASH}
 \sbox{\PARTIALSLASH}{$\partial$\hspace{-1.6mm}/}
 
 \newsavebox{\ASLASH}
 \sbox{\ASLASH}{$A$\hspace{-2.1mm}/}
 
 \newsavebox{\KSLASH}
 \sbox{\KSLASH}{$k$\hspace{-1.8mm}/}
 
 \newsavebox{\LSLASH}
 \sbox{\LSLASH}{$\ell$\hspace{-1.8mm}/}
 
 \newsavebox{\QSLASH}
 \sbox{\QSLASH}{$q$\hspace{-1.8mm}/}
 
 \newsavebox{\DSLASH}
 \sbox{\DSLASH}{$D$\hspace{-2.2mm}/}
 
 \newsavebox{\DbfSLASH}
 \sbox{\DbfSLASH}{${\mathbf D}$\hspace{-2.8mm}/}
 
 \newsavebox{\DELVECRIGHT}
 \sbox{\DELVECRIGHT}{$\stackrel{\rightarrow}{\partial}$}
 
 \newcommand{\blue}{\IfColor{\textCadetBlue}{}}
\newcommand{\black}{\IfColor{\textBlack}{}}
\newcommand{\red}{\IfColor{\textRed}{}}
\newcommand{\green}{\IfColor{\textOliveGreen}{}}
\newcommand{\lila}{\IfColor{\textRedViolet}{}}

\begin{document}
\begin{flushright}
 [math-ph]
\end{flushright}
\title{Geometry and Topology of Anti-BRST Symmetry\\in\\ Quantized Yang-Mills Gauge Theories}

\author{Amir Abbass Varshovi}\email{ab.varshovi@sci.ui.ac.ir/amirabbassv@ipm.ir/amirabbassv@gmail.com}

\affiliation{Faculty of Mathematics and Statistics, Department of Applied Mathematics
and Computer Science, University of Isfahan, Isfahan, IRAN.\\
School of Mathematics, Institute for Research in Fundamental
Sciences (IPM), P.O. Box: 19395-5746, Tehran, IRAN.}
\begin{abstract}
\textbf{Abstract\textbf{:}} The entire geometric formulations of the BRST and the anti-BRST structures are worked out in presence of the Nakanishi-Lautrup field. It is shown that in the general form of gauge fixing mechanisms within the Faddeev-Popov quantization approach, the anti-BRST invariance reflects thoroughly the classical symmetry of the Yang-Mills theories with respect to gauge fixing methods. The Nakanishi-Lautrup field is also defined and worked out as a geometric object. This formulation helps us to introduce two absolutely new topological invariants of quantized Yang-Mills theories, so called the \emph{Nakanishi-Lautrup invariants}. The cohomological structure of the anti-BRST symmetry is also studied and the \emph{anti-BRST topological index} is derived accordingly.
\end{abstract}
\pacs{} \maketitle
%%%%%%%%%%%%%%%%%%%%%%%%
\section{Introduction}
\label{introduction}
\indent Soon after Faddeev and Popov worked out their own quantized Lagrangian via the path-integral formulation of Yang-Mills theories \cite{Faddeev-Popov}, it was seen that the derived theory is surprisingly subject to two different symmetries in the quantized setting. But, although only one of these symmetries, the BRST \cite{BRS, T}, was well understood as the quantized version of the classical gauge invariance \cite{Bonora0,A-J, S, F, Voronin, K}, the other one, the so called anti-BRST \cite{Curci, Ojima}, has remained somehow mysterious so far as an accidentally emergent symmetry of the quantized Lagrangian \cite{Henneaux1, Henneaux2, Weinberg}. Although, this symmetry turned out to be adequate for implementing definite constrains within the regularization of the theory, its physical significance and geometric origin is still unknown \cite{Henneaux1, Henneaux2, Weinberg, Bertlmann}.

\par Actually, despite the BRST symmetry which is well understood from the geometric viewpoints, there is still no admitted idea for such an underlying source for the anti-BRST.\footnote{See \cite{Bonora-super, Robert, Henneaux1, Henneaux2, Henneaux3, Henneaux4, Henneaux5, Henneaux6, Henneaux7, Hull, Barnich, Upadhyay, Bonora1, Bonora2, Varshovi} for different proposals of geometric formulation of anti-BRST symmetry.} Moreover, it seems that as far as we could not provide such an appropriate setting for the anti-BRST structure, it keeps its mystical appearance in quantized gauge theories. But, there is a conflict within the formulation of the anti-BRST invariance. On the one hand, an admirable geometric formulation of the anti-BRST symmetry, must essentially be worked out from a symmetric structure of the classical theory. But, on the other hand, the anti-BRST invariance is actually understood as an extra quantum symmetry with no classical counterpart \cite{Alvarez-Gaume}.

\par Despite this misunderstanding we have already shown that in some special cases of vanishing the Nakanishi-Lautrup auxiliary field \cite{Nakanishi, Lautrup}, the anti-BRST invariance can be figured out within a beautiful geometric picture as the quantized version of an infinite dimensional classical symmetry due to the gauge fixing invariance of the underlying classical gauge theory. In this article we follow the procedure of our model in \cite{Varshovi1} to work out the full BRST-anti-BRST algebra in its most general setting.\footnote{See also \cite{resh} which considers a generalized version of the Faddeev-Popov method, but through with a distinct approach. Also in \cite{Dai} a more generalized formalism via the Batalin-Vilkoviski approach is considered, which is comparable to our model being introduced here.} In fact, we will establish that the invariance of a Yang-Mills gauge theory against different gauge fixing procedures is the classical version of the quantum anti-BRST symmetry. It is obviously an infinite dimensional symmetry even in the classical formulation of Yang-Mills theories. The aspects of this symmetry, just similar to that of the BRST, appears in the quantized Lagrangian only after the gauge is fixed in the Faddeev-Popov path-integral.

\par The main idea to perform the promising project is to study the variations of the Nakanishi-Lautrup field and its regularization role in the framework of the Faddeev-Popov path-integral formula. In fact, since the Nakanishi-Lautrup field has no dynamics in its nature it automatically switches to gauge fixing function $\frak F$ through with the path-integration. Therefore, it seems that the variation of the Nakanishi-Lautrup field plays the role of the variation of $\frak F$ within the Faddeev-Popov quantized Lagrangian. This fact can be considered as the most significant clue to discover the geometric meaning of the anti-BRST structure.

\par As we have already shown via a geometric procedure \cite{Varshovi1}, although incorporating the gauge fixing term in the quantized Lagrangian causes a restriction of the vastly huge domain of the Faddeev-Popov path-integral to a quite smaller space of roots of the gauge fixing function $\frak F$, it provides a set of new transformations to the quantum fields due to continuous variations of $\frak F$.\footnote{This has been proved in the case of sharp or strict gauge fixing methods which correspond to vanishing the Nakanishi-Lautrup field.} This set of new transformations is generalized and established here, as the anti-BRST transformations for any arbitrary gauge fixing method. Therefore, as a final project, this article establishes the correlation of the quantum anti-BRST symmetry and the classical gauge fixing invariance via the Faddeev-Popov quantization approach. However, since the classical theory is physically consistent, we expect that the quantized theory remains invariant against the anti-BRST transformations.

\par In the next section we will give a brief summary of our model in \cite{Varshovi1}.\footnote{We believe that this reviewing section is necessary, because the final project is actually based on the elaborated geometric structures and the constructed definitions we introduced in \cite{Varshovi1}.} Then, in section III, we will derive the general geometric formula of the anti-BRST symmetry and the Nakanishi-Lautrup field by considering the smooth gauge fixing mechanisms. Finally, in the last section, we will have a brief review on the topological features of the anti-BRST symmetry via the given geometric model. We will extract three topological invariants of the quantized Yang-Mills theories based on the anti-BRST structure.\footnote{Although there may be various similar settings among all gauge theories such as the Yang-Mills, $p$-form gauge theories, the gravity, that of the strings, ..., most of our elaborations here are specified to the Yang-Mills gauge theories.}

%%%%%%%%%%%%%%%%%%%%%%%%%%%%%%%%%%%%%%%%

\par
\section{Reviewing The Method: A Geometric Picture of the Anti-BRST Structure}
\setcounter{equation}{0}

\par Let $(\mathcal P,\pi,\mathcal X)$ be a principal $\mathcal G$-bundle for $\mathcal G$ a compact Lie group with Lie algebra $\frak{g}$. The quotient space $\mathcal X=\mathcal P/\mathcal G$ is here understood as the spacetime manifold.\footnote{The spacetime $\mathcal X$ can simply be assumed as $\mathbb{R}^n$ or $S^n=\mathbb{R}^n \cup \{\infty\}$. However, for simplicity, one may consider $\mathcal P$ to be topologically trivial, i.e. $\mathcal P=\mathcal X \times \mathcal G$. Although, our framework is worked out for general topological structures.} The space of Cartan connection forms over $\mathcal P$, so called $\widetilde{\mathcal P}$, is subject to the action of smooth maps $\widetilde{g}:\mathcal X\to \mathcal G$ from right;\footnote{Here, $\emph{\emph{d}}$ is the exterior derivative operator of $\mathcal P$. Moreover, here $\widetilde{g} \circ \pi$ has been replaced with $\widetilde{g}$ for simplicity.}
\begin{eqnarray} \label {1} 
~~~~~~~~~~~~~~~~~~~~\widetilde{\sigma}.\widetilde{g}=\widetilde{g}^{-1} \emph{\emph{d}}\widetilde{g}+\text{Ad}_{\widetilde{g}^{-1}} (\widetilde{\sigma}) ,~~~~~~~~~~(\widetilde{\sigma} \in \widetilde{\mathcal P}).
\end{eqnarray}

\par Let $\widetilde{\mathcal G}$ be the space of fixed point preserving maps $\widetilde{g}:\mathcal X\to \mathcal G$, so called the \emph{gauge transformations}.\footnote{See \cite{Varshovi1} for more details.} Thus, for each $x \in \mathcal X$, there is a projective group homomorphism $\phi_x:\widetilde{\mathcal G}\to \mathcal G$, with $\phi_x(\widetilde{g})=\widetilde{g}(x)$.\footnote{This condition is fulfilled except at the infinity point of $S^n=\mathbb{R}^n \cup \{\infty\}$ \cite{Varshovi1}.} The \emph{infinitesimal gauge transformations}, i.e. the smooth maps $\widetilde{\alpha}:\mathcal X\to \frak{g}$, produce the Lie algebra of $\widetilde{\mathcal G}$. We denote it by $\widetilde{\frak{g}}$.

\par Set  $\widetilde{\mathcal X}=\widetilde{\mathcal P}/\widetilde{\mathcal G}$. Hence, similarly, the triplet of $(\widetilde{\mathcal P},\widetilde{\pi},\widetilde{\mathcal X})$ provides a principal $\widetilde{\mathcal G}$-bundle. The topology of this principal bundle is not trivial even if $(\mathcal P,\pi,\mathcal X)$ is so \cite{Singer}.\footnote{Actually, this is true for non-Abelian groups $SU(N)$, $N \ge 2$. It is mostly called the Gribov ambiguity \cite{Gribov}.} Now let $\widetilde{\widetilde{\mathcal P}}$ be the space of parallelism structures on $\widetilde{\mathcal P}$. Actually, $\widetilde{\widetilde{\mathcal P}}$ is also foliated modulo adjoint transformations which are performed by action of smooth $\widetilde{\widetilde{g}}:\widetilde{\mathcal X}\to \widetilde{\mathcal G}$ as;
\begin{eqnarray} \label {2} 
~~~~~~~~~~~~~~~~~~~~\widetilde{\widetilde{\sigma}}.\widetilde{\widetilde{g}}=\widetilde{\widetilde{g}}^{-1} \tilde{\emph{\emph{d}}} \widetilde{\widetilde{g}}+\text{Ad}_{\widetilde{\widetilde{g}}^{-1}} (\widetilde{\widetilde{\sigma}}) ,~~~~~~~~~~(\widetilde{\widetilde{\sigma}} \in \widetilde{\widetilde{\mathcal P}}),
\end{eqnarray}

\noindent wherein $\tilde{\emph{\emph{d}}}$ is the exterior derivative operator of $\widetilde{\mathcal P}$.\footnote{We used $\widetilde{\widetilde{g}}$ instead of $\widetilde{\widetilde{g}} \circ \widetilde{\pi}$.} The space of all $\widetilde{\widetilde{g}}$s in (\ref{2}), shown with $\widetilde{\widetilde{\mathcal G}}$, is in fact a Lie group with a fairly hidden role in Yang-Mills theories. In principle, its elements transform (local) sections of the principal bundle $\widetilde{\mathcal P} \to \widetilde{\mathcal X}$ to each other. Hence, since each of these (local) sections displays a gauge fixing mechanism in the Yang-Mills theory, the elements of $\widetilde{\widetilde{\mathcal G}}$ are referred to as \emph{gauge fixing transformations}.  Consequently, $\widetilde{\widetilde{\mathcal G}}$ itself is called the \emph{gauge fixing transformation group}. Once again, for each $\widetilde{x}\in \widetilde{\mathcal X}$, we obtain a projective group homomorphism as $\widetilde{\phi}_{\widetilde{x}}:\widetilde{\widetilde{\mathcal G}}\to \widetilde{\mathcal G}$, with $\widetilde{\phi}_{\widetilde{x}}(\widetilde{\widetilde{g}})=\widetilde{\widetilde{g}}(\widetilde{x})$. Shown with $\widetilde{\widetilde{\frak{g}}}$, the space of the generators of $\widetilde{\widetilde{\mathcal G}}$ (the Lie algebra of $\widetilde{\widetilde{\mathcal G}}$), consists of smooth maps $\widetilde{\widetilde{\alpha}}:\widetilde{\mathcal X}\to \widetilde{\frak{g}}$, each of which is referred to as an \emph{infinitesimal gauge fixing transformation}.

\par To figure out the significance of the above structures one must first define the following map;
\begin{eqnarray} \label {3}
\begin{array}{*{20}{c}}
\frak i:\widetilde{\mathcal P}\times \widetilde{\mathcal G}\times \widetilde{\widetilde{\mathcal G}}\to \widetilde{\mathcal P}~~~~~~~~~~~~~~~ \\
~~~~~~~~~~~~~~(\widetilde{\sigma},\widetilde{g},\widetilde{\widetilde{g}})~~~\mapsto \widetilde{\sigma}.\widetilde{g}.\phi_{\widetilde{x}}(\widetilde{\widetilde{g}})=\widetilde{\sigma}.\widetilde{g}\widetilde{\widetilde{g}}(\widetilde{x}), \\
\end{array}
\end{eqnarray}

\noindent with $\widetilde{x}=\widetilde{\pi}(\widetilde{\sigma})\in \widetilde{X}$. Now, assume a Cartan connection form on $\widetilde{\mathcal P}$, say $\widetilde{\widetilde{\sigma}}_0 \in \widetilde{\widetilde{\mathcal P}}$, and let $\Pi$ be its pull-back through $\frak i$ as a differential one-form over $\widetilde{\mathcal P}\times \widetilde{\mathcal G}\times \widetilde{\widetilde{\mathcal G}}$ with values in $\widetilde{\frak{g}}$. We read;
\begin{eqnarray} \label {5}
\Pi_{(\widetilde{\sigma},\widetilde{g},\widetilde{\widetilde{g}})}=\frak i^*(\widetilde{\widetilde{\sigma}}_0)_{(\widetilde{\sigma},\widetilde{g},\widetilde{\widetilde{g}})}=\text{Ad}_{\widetilde{\widetilde{g}} (\widetilde{x})^{-1}\widetilde{g}^{-1}} (\widetilde{\widetilde{\sigma}}_0)_{\widetilde{\sigma}}+\text{Ad}_{\widetilde{\widetilde{g}}(\widetilde{x})^{-1}}(\widetilde{\omega}_{\widetilde{g}})+\widetilde{\widetilde{\omega}}_{\widetilde{\widetilde{g}}}(\widetilde{x}),
\end{eqnarray}

\noindent wherein $\widetilde{\omega}$ and $\widetilde{\widetilde{\omega}}$ are respectively the Maurer-Cartan forms on $\widetilde{\mathcal G}$ and $\widetilde{\widetilde{\mathcal G}}$. We define $\Omega$ as;
$$\Omega:T\mathcal P\times T\widetilde{\mathcal P}\times T\widetilde{\mathcal G}\times T\widetilde{\widetilde{\mathcal G}} \to \frak{g}$$
\begin{eqnarray} \label {6}
\Omega((v,\widetilde{v},\widetilde{\alpha},\widetilde{\widetilde{\alpha}})_{(\sigma,\widetilde{\sigma},\widetilde{g},\widetilde{\widetilde{g}})})=\widetilde{\sigma}.\widetilde{g} \widetilde{\widetilde{g}} (\widetilde{x})(v_\sigma)+\Pi((\widetilde{v},\widetilde{\alpha},\widetilde{\widetilde{\alpha}})_{(\widetilde{\sigma},\widetilde{g},\widetilde{\widetilde{g}})})(x),
\end{eqnarray}

\noindent for $\widetilde{x}=\widetilde{\pi}(\widetilde{\sigma})$ and $x=\pi (\sigma)$. Actually, $\Omega$ is a $\frak{g}$-valued one-form over $\mathcal P\times\widetilde{\mathcal P}\times \widetilde{\mathcal G}\times \widetilde{\widetilde{\mathcal G}}$. As we mentioned above, any section of $(\widetilde{\mathcal P},\widetilde{\pi},\widetilde{\mathcal X})$, is in principle a gauge fixing function, a geometric machinery which selects a specific gauge field amongst all of its equivalent forms via the gauge transformation process.\footnote{In fact, without loss of generality, one can suppose that $(\widetilde{\mathcal P},\widetilde{\pi},\widetilde{\mathcal X})$ admits a densely defined trivialization structure \cite{Varshovi1, Singer}. Hence, sections are assumed to be smooth. For more details see \cite{Varshovi1}.}

\par Set $\mathcal M=\mathcal X\times \widetilde{\mathcal G}\times \widetilde{\widetilde{\mathcal G}}$, consider two sections of $(\mathcal P, \pi, \mathcal X)$ and $(\widetilde{\mathcal P},\widetilde{\pi},\widetilde{\mathcal X})$, respectively shown with $\frak s: \mathcal X \to \mathcal P$ and $\widetilde{\frak s}:\widetilde{\mathcal X}\to \widetilde{\mathcal P}$, and fix some $\widetilde{\sigma}_0\in \widetilde{\mathcal P}$ with $\widetilde{\pi}(\widetilde{\sigma}_0)=\widetilde{x}_0 \in \mathcal X$. Then, we define;
\begin{eqnarray} \label {7}
\begin{array}{*{20}{c}}
~~~~~~~~~~~~~ \frak j_{\frak s,\widetilde{\frak s},\widetilde {\sigma}_0}:~~\mathcal M ~~~\to ~~~ \mathcal P\times\widetilde{\mathcal P}\times \widetilde{\mathcal G}\times \widetilde{\widetilde{\mathcal G}}~~~~~~~~~~~~~~~ \\
~~~~~~~(x,\widetilde{g},\widetilde{\widetilde{g}})~\mapsto ~(\frak s(x),\widetilde{\frak s}(\widetilde{x}_0)),\widetilde{g},\widetilde{\widetilde{g}}). \\
\end{array}
\end{eqnarray}

\noindent The pull-back of $\Omega$ through $\frak j_{\frak s,\widetilde{\frak s},\widetilde {\sigma}_0}$ is a $\frak{g}$-valued one-form over $\mathcal M$ as;
\begin{eqnarray} \label {8}
\frak j_{\frak s,\widetilde{\frak s},\widetilde {\sigma}_0}^*(\Omega)_{(x,\widetilde{g},\widetilde{\widetilde{g}})}=\mathcal A_{(x,\widetilde{g},\widetilde{\widetilde{g}})}+\mathcal C_{(x,\widetilde{g},\widetilde{\widetilde{g}})}+\widetilde{\mathcal C}_{(x,\widetilde{g},\widetilde{\widetilde{g}})},
\end{eqnarray}

\noindent where;
\begin{eqnarray} \label {9}
\mathcal A_{(x,\widetilde{g},\widetilde{\widetilde{g}})}=\frak s^*(\widetilde{s}(\widetilde{x}_0).\widetilde{g}\widetilde{\widetilde{g}}(\widetilde{x}_0))_x,~~~~\mathcal C_{(x,\widetilde{g},\widetilde{\widetilde{g}})}=\text{Ad}_{\widetilde{\widetilde{g}}(\widetilde{x}_0)^{-1}}(\widetilde{\omega}_{\widetilde{g}})(x),~~~~~\widetilde{\mathcal C}_{(x,\widetilde{g},\widetilde{\widetilde{g}})}=\widetilde{\widetilde{\omega}}_{\widetilde{\widetilde{g}}}(\widetilde{x}_0)(x),
\end{eqnarray}
\noindent are respectively called \emph{gauge}, \emph{ghost} and \emph{anti-ghost} fields. The exterior derivative of $\mathcal M$, naturally splits to three distinct parts for components $\mathcal X$, $\widetilde{\mathcal G}$ and $\widetilde{\widetilde{\mathcal G}}$, respectively, i.e. $\emph{\emph{d}}_{\mathcal M}=\emph{\emph{d}}+\delta+\widetilde{\delta}$, for $\emph{\emph{d}}=\text{d}_{\mathcal X}$, $\delta=\text{d}_{\mathcal G}$, and $\widetilde{\delta}=\text{d}_{\widetilde {\mathcal G}}$. Thus, we find the following immediate formulas;
\begin{eqnarray} \label {11}
\delta \mathcal A=-\emph{\emph{d}}\mathcal C-\mathcal A \mathcal C-\mathcal C \mathcal A,~~~~~ \delta \mathcal C=-\mathcal C^2,~~~~~\delta \widetilde{\mathcal C}=0,~
\end{eqnarray}

\noindent accompanied with;
\begin{eqnarray} \label {12}
\widetilde{\delta} \mathcal A=-\emph{\emph{d}}\widetilde{\mathcal C}-\mathcal A\widetilde{\mathcal C}-\widetilde{\mathcal C}\mathcal A,~~~~~\widetilde{\delta} \mathcal C=-\mathcal C\widetilde{\mathcal C}-\widetilde{\mathcal C}\mathcal C,~~~~~\widetilde{\delta} \widetilde{\mathcal C}=-\widetilde{\mathcal C}^2.
\end{eqnarray}

\noindent On the other hand, since $\text{d}_{\mathcal G}$, $\text{d}_{\widetilde{\mathcal G}}$ and $\emph{\emph{d}}_{\mathcal M}$ are nilpotent operators one readily obtains;
\begin{eqnarray} \label {13}
\delta^2=0,~~~~~~~~~~\widetilde{\delta}^2=0,~~~~~~~~~~\emph{\emph{d}}\delta+\delta\emph{\emph{d}}=0,~~~~~~~~~~\emph{\emph{d}}\widetilde{\delta}+\widetilde{\delta}\emph{\emph{d}}=0,~~~~~~~~~~\delta\widetilde{\delta}+\widetilde{\delta}\delta=0.
\end{eqnarray}

\noindent However, for vanishing Nakanishi-Lautrup field, (\ref {11}) to (\ref {13}) are in complete agreement with the well-known formulas of the BRST and the anti-BRST derivatives, considered as operators $\delta$ and $\widetilde{\delta}$ accordingly. Thus, from now on we refer to $\delta$ (resp. $\widetilde{\delta}$) as the \emph{BRST operator} (derivative) (resp. the \emph{anti-BRST operator} (derivative)).

\par Contributing $\widetilde{\widetilde{\mathcal G}}$ in the Faddeev-Popov Lagrangian via the anti-ghost field is reflected by different gauge fixing mechanisms. Thus, based on its essence which gives no address to gauge fixing procedures in its classical formalism, a Yang-Mills gauge theory, is expected to be symmetric under the action of the gauge fixing transformation group $\widetilde{\widetilde{\mathcal G}}$ at all quantum levels. Therefore, as a smooth function over $\mathcal M$, the Faddeev-Popov Lagrangian ought to be a closed form for anti-BRST derivative $\widetilde{\delta}$. In fact, This is the geometric explanation of the anti-BRST invariance of the Faddeev-Popov quantized Lagrangian \cite {Weinberg}.\footnote{See \cite {Alvarez-Gaume} for more discussions about the anti-BRST symmetry of the Faddeev-Popov Lagrangian.} Thus, we obtain the following theorem;\\

\textbf{Theorem 1\textbf{;} \cite{Varshovi1}} \emph{In any quantized Yang-Mills gauge theory with a strict gauge fixing mechanism (vanishing Nakanishi-Lautrup field) the anti-BRST invariance is the quantized version of classical gauge fixing symmetry. Moreover, $\mathcal A$, $\mathcal C$ and $\widetilde{\mathcal C}$ (introduced in (\ref {9}) as the gauge, the ghost and the anti-ghost fields), together with the exterior derivative operators of Lie groups $\widetilde{\mathcal G}$ and $\widetilde{\widetilde{\mathcal G}}$, respectively shown by $\delta$ and $\widetilde{\delta}$ (and considered as the BRST and the anti-BRST derivatives), produce the BRST-anti-BRST algebra with vanishing Nakanishi-Lautrup field (i.e. for strict gauge fixing mechanisms).}\\

%%%%%%%%%%%%%%%%%%%%%%%%%%%%%%%%%%%%%%%%%%%%%%%%%%%%%

\par
\section{Final Project: Anti-BRST Symmetry in Presence of Nakanishi-Lautrup Field}
\setcounter{equation}{0}
\par Despite of its interesting results, the \textbf{Theorem 1} suffers from a technical failure due to its limits, to be true only for strict gauges or vanishing Nakanishi-Lautrup field. In order to work out the general formulation of the anti-BRST structure in the Faddeev-Popov Lagrangian one needs to modify the definition of the ghost/anti-ghost fields of the previous section in an appropriate setting. We should recall that the sharp gauge fixing mechanisms are particularly imposed by incorporating Dirac delta functions in the path-integral formulation. In contrast to simplicity of such mechanisms, they don't admit smooth enough structures to be modeled via differential geometric tools in an appropriate manner. Therefore, one may naturally be interested in smooth gauge fixings.

\par Before going further let us fix some conventions in our terminology. We have two seemingly different meaning for the \emph{gauge fixing function}. The first is an algebro-geometric notion which is implemented by $\widetilde{\frak{g}}$-valued functions over $\widetilde{\mathcal P}$. Here, we note that if $\widetilde {\frak F}$ is an appropriate smooth $\widetilde{\frak{g}}$-valued function over $\widetilde{\mathcal P}$ its root defines a smooth section of $(\widetilde{\mathcal P},\widetilde{\pi},\widetilde{\mathcal X})$, say $\widetilde{\frak s}$, which is in fact, the second notion for the gauge fixing function in our formalism. More strictly, if for each $\widetilde x \in \widetilde{\mathcal X}$, the smooth function $\widetilde {\frak F}$ vanishes at a unique element of $\widetilde{\pi}^{-1}(\widetilde{x})$, say $\widetilde {\frak s} (\widetilde x)$, then the corresponding  map of $\widetilde{\frak s}:\widetilde{\mathcal X} \to \widetilde{\mathcal P}$, with $\widetilde{x} \mapsto \widetilde{\frak s}(\widetilde{x})$, provides a smooth section of $\widetilde{\pi}$. We refer to both $\frak F$ and $\widetilde{\frak s}$ as the gauge fixing function and we will keep in mind their correspondence.

\par But, however, in smooth approaches of fixing the gauge a $\widetilde{\frak{g}}$-valued function over $\widetilde{\mathcal P}$, say $\frak F$, is incorporated into the path-integral, mostly by means of a Gaussian functional, to confine the path-integral domain to those gauge fields which almost obey the gauge fixing laws imposed by $\frak F$, i.e. the roots of $\frak F$ and their nearby elements in $\widetilde{\mathcal P}$ \cite{Weinberg}. To impose such gauge fixing mechanisms to path-integral formulation of Yang-Mills theories one should revise the constructed geometric structures in the previous section accordingly. At the first step we note that the gauge fixing term $\frak F$ provides a smooth map from $\widetilde{\mathcal G}$ to $\widetilde{\widetilde{\mathcal G}}$ as;
\begin{eqnarray} \label {2-1}
\begin{array}{*{20}{c}}
~~~~~ \theta:\widetilde{\mathcal G}\to \widetilde{\widetilde{\mathcal G}}~~~~~~~~~~~~~~~ \\
~~~~~~~~ \widetilde{g}\mapsto \exp (\frak F(\widetilde{\frak s}.\widetilde{g})), \\
\end{array}
\end{eqnarray}

\noindent wherein $\widetilde{\frak s}:\widetilde{\mathcal X} \to \widetilde{\mathcal P}$ is the root of $\frak F$. Actually, as we see in below, $\theta$ extends the domain of the Faddeev-Popov path-integral to contain smoothly varying gauges defined by the gauge fixing transformation $\theta(\widetilde g)$. More precisely, if $\frak s':\widetilde{\mathcal X} \to \widetilde{\mathcal P}$ is a (local) section in the neighborhood of $\frak s$, defined with $\frak s'=\frak s.\widetilde g$ for some $\widetilde g\in \widetilde {\mathcal G}$, then it is assumed to be subject to the $\theta(\widetilde{g})$-transformation of $\frak F$, i.e. $\frak F_{\theta}(\widetilde{\sigma})=\frak F(\widetilde{\sigma}.\widetilde{g})$. As we show in the following this process will revive the Nakanishi-Lautrup field in our formalism to include the smoothly defined gauge fixing mechanisms.

\par We replace $\frak i$ in (\ref {3}) by $\frak i_{\theta}$, defined as:
\begin{eqnarray} \label {2-2}
\begin{array}{*{20}{c}}
\frak i_{\theta}:\widetilde{\mathcal P}\times \widetilde{\mathcal G}\times \widetilde{\widetilde{\mathcal G}}\to \widetilde{\mathcal P}~~~~~~~~~~~~~~~ \\
~~~~~~~(\widetilde{\sigma},\widetilde{g},\widetilde{\widetilde{g}})~~~\mapsto \widetilde{\sigma}.\widetilde{g}.(\widetilde{\widetilde{g}}\theta(\widetilde{g}))(\widetilde{x}), \\
\end{array}
\end{eqnarray}

\noindent for $\widetilde{x}=\widetilde{\pi}(\widetilde{\sigma})$. Similarly, we obtain;
\begin{eqnarray} \label {2-3}
\begin{array}{*{20}{c}}
~~~~~ {\Pi_{\theta}}_{(\widetilde{\sigma},\widetilde{g},\widetilde{\widetilde{g}})}=\frak i_{\theta}^*(\widetilde{\widetilde{\sigma}}_0)_{(\widetilde{\sigma},\widetilde{g},\widetilde{\widetilde{g}})}=~~~~~~~~~~~~~~~ \\
\text{Ad}_{(\widetilde{\widetilde{g}}\theta(\widetilde{g})) (\widetilde{x})^{-1}\widetilde{g}^{-1}} (\widetilde{\widetilde{\sigma}}_0)_{\widetilde{\sigma}}+(\theta(\widetilde{g})^{-1} \text{d}_{\widetilde{\mathcal G}}\theta(\widetilde{g}))_{\widetilde{g}}(\widetilde{x})+\text{Ad}_{(\widetilde{\widetilde{g}}\theta(\widetilde{g}))(\widetilde{x})^{-1}}(\widetilde{\omega}_{\widetilde{g}})
+\text{Ad}_{\theta(\widetilde{g})^{-1}}(\widetilde{\widetilde{\omega}}_{\widetilde{\widetilde{g}}})(\widetilde{x}), \\
\end{array}
\end{eqnarray}

\noindent for $\widetilde{\widetilde{\sigma}}_0$ a fixed chosen connection over $\widetilde{\mathcal P}$.\footnote{It is worth to note that for each $\frak h \in C^\infty(\widetilde {\mathcal G})$ we have $\emph{\emph{d}}_{\widetilde{\mathcal G}}\frak{h}=\frac{{\delta }}{{\delta \widetilde{g} }}\frak{h}.\widetilde{\omega}$, where $\widetilde{\omega}$ is the Cartan form over $\widetilde{\mathcal G}$. That is; $\theta(\widetilde{g})^{-1}\emph{\emph{d}}_{\widetilde{\mathcal G}}\theta(\widetilde{g})=\frac{{\delta }}{{\delta \widetilde{g}^a }}\frak F(\sigma.\widetilde{g})\widetilde{\omega}^a$ for $a$ the index of the generators of $\mathcal G$. Actually, $\widetilde{\mathcal C}^b \frac{{\delta }}{{\delta \widetilde{g}^a }}\frak F^b(\sigma .\widetilde{g})\mathcal C^a$ appears explicitly in the Faddeev-Popov Lagrangian as the gauge fixing term.} Thus, here $\Pi_{\theta}$ is once again a $\widetilde{\frak{g}}$-valued one-form over $\widetilde{\mathcal P}\times \widetilde{\mathcal G}\times \widetilde{\widetilde{\mathcal G}}$. Next we define a $\frak{g}$-valued one-form over $\mathcal P\times\widetilde{\mathcal P}\times \widetilde{\mathcal G}\times \widetilde{\widetilde{\mathcal G}}$, say $\Omega_{\theta}$, with
\begin{eqnarray} \label {2-4}
\Omega_{\theta}(v,\widetilde{v},\widetilde{\alpha},\widetilde{\widetilde{\alpha}})
=\widetilde{\sigma}.\widetilde{g}.(\widetilde{\widetilde{g}}\theta({\widetilde{g}})) (\widetilde{x})(v )+\Pi_{\theta} ((\widetilde{v},\widetilde{\alpha},\widetilde{\widetilde{\alpha}})(x),
\end{eqnarray}

\noindent for $(v,\widetilde{v},\widetilde{\alpha},\widetilde{\widetilde{\alpha}}) \in T_{\sigma} \mathcal P \times T_{\widetilde{\sigma}} \widetilde{\mathcal P} \times T_{\widetilde{g}} \widetilde{\mathcal G} \times T_{\widetilde{\widetilde{g}}}\widetilde{\widetilde{\mathcal G}}$, with $\widetilde{x}=\widetilde{\pi}(\widetilde{\sigma})$ and $x=\pi (\sigma)$.

\par Then, for two given sections $\frak s : \mathcal X \to \mathcal P$ and $\widetilde{\frak s}: \widetilde{\mathcal X} \to \widetilde{\mathcal P}$ (the gauge fixing section), and by fixing an element in $\widetilde{\mathcal P}$, say $\widetilde{\sigma}_0$, with $\widetilde{\pi}(\widetilde{\sigma}_0)=\widetilde{x}_0$, we can similarly define $\frak j_{\frak s, \widetilde{\frak s}, \widetilde{\sigma}_0}$ of (\ref {7}). The pull-back of $\Omega_{\theta}$ through $\frak j_{\frak s, \widetilde{\frak s}, \widetilde{\sigma}_0}$ is a $\frak{g}$-valued one-form over $\mathcal M=\mathcal X\times \widetilde{\mathcal G}\times \widetilde{\widetilde{\mathcal G}}$ with;
\begin{eqnarray} \label {2-5}
\begin{array}{*{20}{c}}
~~~~~ \frak j_{\frak s, \widetilde{\frak s}, \widetilde{\sigma}_0}^*(\Omega_{\theta})_{(x,\widetilde{g},\widetilde{\widetilde{g}})}=\frak s^*(\widetilde{\frak s}(\widetilde{x}_0).\widetilde{g}.(\widetilde{\widetilde{g}}\theta(\widetilde{g}))(\widetilde{x}_0))_x \\
+(\theta(\widetilde{g})^{-1}\emph{\emph{d}}_{\widetilde{\mathcal G}}\theta(\widetilde{g}))_{\widetilde{g}}(\widetilde{x}_0)(x)+\text{Ad}_{(\widetilde{\widetilde{g}}\theta(\widetilde{g}))(\widetilde{x}_0)^{-1}}(\widetilde{\omega}_{\widetilde{g}})(x)+\text{Ad}_{\theta(\widetilde{g})^{-1}}(\widetilde{\widetilde{\omega}}_{\widetilde{\widetilde{g}}})(\widetilde{x}_0)(x), \\
\end{array}
\end{eqnarray}

\noindent Let us redefine the gauge field $\mathcal A$ as the first term of $\frak j_{\frak s, \widetilde{\frak s}, \widetilde{\sigma}_0}^*(\Omega_{\theta})$, i.e.;
\begin{eqnarray} \label {2-6}
\mathcal A_{(x,\widetilde{g},\widetilde{\widetilde{g}})}=\frak s^*(\widetilde{\frak s}(\widetilde{x}_0).\widetilde{g}(\widetilde{\widetilde{g}}\theta(\widetilde{g}))(\widetilde{x}_0))_x,
\end{eqnarray}

\noindent which is in fact a $\frak g$-valued one-form on $\mathcal X$. Next, we redefine the ghost field, similarly shown with $\mathcal C$, with those elements of $\frak j_{\frak s, \widetilde{\frak s}, \widetilde{\sigma}_0}^*(\Omega_{\theta})$ which are one-forms over $\widetilde{\mathcal G}$. Thus, $\mathcal C$ is actually equal to the sum of the second and third terms of the right hand side of (\ref {2-5});
\begin{eqnarray} \label {2-7}
\mathcal C_{(x,\widetilde{g},\widetilde{\widetilde{g}})}=(\theta(\widetilde{g})^{-1}\emph{\emph{d}}_{\widetilde{\mathcal G}}\theta(\widetilde{g}))_{\widetilde{g}}(\widetilde{x}_0)(x)+\text{Ad}_{(\widetilde{\widetilde{g}}\theta(\widetilde{g}))(\widetilde{x}_0)^{-1}}(\widetilde{\omega}_{\widetilde{g}})(x).
\end{eqnarray}

\noindent For some more precisions in our next manipulations let us show the first and the second parts of $\mathcal C$ respectively with $\mathcal B$ and $\mathcal B'$, i.e. $\mathcal C=\mathcal B+\mathcal B'$, with;
\begin{equation} \label {bb'}
\mathcal B_{(x,\widetilde{g},\widetilde{\widetilde{g}})}=(\theta(\widetilde{g})^{-1}\emph{\emph{d}}_{\widetilde{\mathcal G}}\theta(\widetilde{g}))_{\widetilde{g}}(\widetilde{x}_0)(x),~~~~~~~~~~\mathcal B'_{(x,\widetilde{g},\widetilde{\widetilde{g}})}=\text{Ad}_{(\widetilde{\widetilde{g}}\theta(\widetilde{g}))(\widetilde{x}_0)^{-1}}(\widetilde{\omega}_{\widetilde{g}})(x).
\end{equation}

\noindent The anti-ghost field $\widetilde{\mathcal C}$ is the only part of $\frak j_{\frak s, \widetilde{\frak s}, \widetilde{\sigma}_0}^*(\Omega_{\theta})$ which defines a one-form on $\widetilde{\widetilde{\mathcal G}}$, i.e. the last term of the right hand side of (\ref{2-5});
\begin{eqnarray} \label {2-8}
\widetilde{\mathcal C}_{(x,\widetilde{g},\widetilde{\widetilde{g}})}=\text{Ad}_{\theta(\widetilde{g})^{-1}}(\widetilde{\widetilde{\omega}}_{\widetilde{\widetilde{g}}})(\widetilde{x}_0)(x).
\end{eqnarray}

\noindent Similarly, define the BRST and anti-BRST derivatives as the exterior derivative operators of $\widetilde{\mathcal G}$ and of $\widetilde{\widetilde{G}}$, i.e. $\delta=\text{d}_{\widetilde{\mathcal G}}$ and $\widetilde {\delta}=\text{d}_{\widetilde{\widetilde{\mathcal G}}}$. It can be easily seen that;
\begin{eqnarray} \label {2-9}
\delta \mathcal B=-\mathcal B^2,~~~~~\delta \mathcal B'=-\mathcal B\mathcal B'-\mathcal B'\mathcal B-{\mathcal B}'^2.
\end{eqnarray}
\noindent These equations naturally lead to;
\begin{eqnarray} \label {2-10}
\delta \mathcal C=-\mathcal C^2.
\end{eqnarray}

\noindent Moreover, it is easy to see that;
\begin{equation} \label{delta-anti-ghost}
\delta \widetilde{\mathcal C}=-\mathcal B \widetilde{\mathcal C}-\widetilde{\mathcal C} \mathcal B.
\end{equation}

\noindent On the other hand, upon the definition, the Nakanishi-Lautrup auxiliary field $\mathcal V$ is equal to $-\delta\widetilde{C}$ \cite{Weinberg}, hence we readily obtain from (\ref{delta-anti-ghost});
\begin{eqnarray} \label {2-11}
\mathcal V=\mathcal B \widetilde{\mathcal C}+ \widetilde{\mathcal C}\mathcal B.
\end{eqnarray}

\noindent Obviously, $\mathcal V$ has to be a BRST closed form, i.e.;
\begin{equation} \label{delta-v}
\delta \mathcal V=0,
\end{equation}

\noindent since $\delta$ is a nilpotent operator and $\mathcal V$ is itself a BRST exact form. However, this result could be checked directly using (\ref {2-9}) and (\ref {2-11}).
\par On the other hand, we immediately read as before;
\begin{eqnarray} \label {2-12}
\delta \mathcal A=-\text{d}\mathcal C-\mathcal A\mathcal C-\mathcal C\mathcal A,~~~~~\widetilde{\delta} \mathcal A=-\text{d}\widetilde{\mathcal C}-\mathcal A\widetilde{\mathcal C}-\widetilde{\mathcal C}\mathcal A.
\end{eqnarray}

\noindent Similarly, we see;
\begin{eqnarray} \label {2-13}
\widetilde{\delta}\widetilde{\mathcal C}=-\widetilde{\mathcal C}^2,
\end{eqnarray}

\noindent and;
\begin{eqnarray} \label {2-14}
\widetilde{\delta} \mathcal B=0,~~~~~\widetilde{\delta} \mathcal B'=-\mathcal B'\widetilde{\mathcal C}-\widetilde{\mathcal C}\mathcal B'.
\end{eqnarray}

\noindent Therefore, according to the decomposition $\mathcal C=\mathcal B+\mathcal B'$ together with (\ref {2-11}) and (\ref {2-14}) we find;
\begin{eqnarray} \label {2-15}
\widetilde{\delta}\mathcal C=-\mathcal C\widetilde{\mathcal C}-\widetilde{\mathcal C}\mathcal C+\mathcal V.
\end{eqnarray}

\noindent Finally, from (\ref{2-11}) and (\ref {2-13}) it is seen that;
\begin{eqnarray} \label {2-16}
\widetilde{\delta} \mathcal V=\mathcal V\widetilde{\mathcal C}-\widetilde{\mathcal C}\mathcal V.
\end{eqnarray}

\par Consequently, from (\ref {2-10}) to (\ref {2-16}) we get collectively the following relations;
\begin{eqnarray} \label {2-17}
\begin{array}{*{20}{c}}
\delta \mathcal A=-\text{d}\mathcal C-\mathcal A\mathcal C-\mathcal C\mathcal A,~~~~~ \delta \mathcal C=-\mathcal C^2,~~~~~~~~~~~~~~~~~ \\
\delta \widetilde{\mathcal C}=-\mathcal V,~~~~~~~~~~~~~~~~~~~~~~\delta \mathcal V=0,~~~~~~~~~~~~~~~~~~~~~ \\
\end{array}
\end{eqnarray}

\noindent and;
\begin{eqnarray} \label {2-18}
\begin{array}{*{20}{c}}
\widetilde{\delta} \mathcal A=-\text{d}\widetilde{\mathcal C}-\mathcal A\widetilde{\mathcal C}-\widetilde{\mathcal C}\mathcal A,~~~~~\widetilde{\delta} \mathcal C=-\mathcal C\widetilde{\mathcal C}-\widetilde{\mathcal C}\mathcal C+\mathcal V,~~~ \\
\widetilde{\delta} \widetilde{\mathcal C}=-\widetilde{\mathcal C}^2,~~~~~~~~~~~~~~~~~~~~~~\widetilde{\delta} \mathcal V=\mathcal V\widetilde{\mathcal C}-\widetilde{\mathcal C}\mathcal V,~~~~~~~~~~ \\
\end{array}
\end{eqnarray}

\noindent which both are in agreement with the original BRST-anti-BRST equations (in presence of the Nakanishi-Lautrup field $\mathcal V$), well-known in the literature \cite{Weinberg}.

\par Similarly, it is seen that through with the above geometric elaborations the quantum fields $\mathcal A$, $\mathcal C$, and $\widetilde{\mathcal C}$, and consequently the Faddeev-Popov Lagrangian, all are given as differential forms over the product space of $\mathcal X \times \widetilde{\mathcal X} \times \widetilde{\mathcal G} \times \widetilde{\widetilde{\mathcal G}}$ (roughly speaking over $\mathcal X \times \widetilde{\mathcal P} \times \widetilde{\widetilde{\mathcal G}}$). In fact, this manifold, beside its last component, is the familiar space on which the Faddeev-Popov path-integral is taken place.\footnote{Note that in this framework the only fixed element (beside $\frak s$) is the gauge fixing term $\widetilde{\frak s}$ which we need as a fundamental tool for the Faddeev-Popov quantization method of Yang-Mills theories. More precisely, the path-integration over $\widetilde{\widetilde{\mathcal G}}$ is actually done by considering $\widetilde{\frak s}$ and assuming a finite Haar measure on $\widetilde{\widetilde{\mathcal G}}$, i.e. $\text{Vol}(\widetilde{\widetilde{\mathcal G}})=\int_{\widetilde{\widetilde{\mathcal G}}}\text{dVol}=1$ \cite{Varshovi1}.} Therefore, we have already established the following theorem; \\

\textbf{Theorem 2\textbf{;}} \emph{In any quantized Yang-Mills gauge theory with general gauge fixing mechanism the anti-BRST invariance is the quantized version of classical gauge fixing symmetry. Moreover, $\mathcal A$, $\mathcal C$, $\widetilde{\mathcal C}$ and $\mathcal V$ introduced in (\ref {2-6}), (\ref {2-7}), (\ref {2-8}) and (\ref {2-11}) respectively as the gauge, the ghost, the anti-ghost and the Nakanishi-Lautrup fields together with exterior derivative operators of $\widetilde{\mathcal G}$ and $\widetilde{\widetilde{\mathcal G}}$, shown accordingly with $\delta$ and $\widetilde{\delta}$, as the BRST and the anti-BRST derivatives, produce the full BRST-anti-BRST algebra of quantized Yang-Mills gauge theories.} \\

%%%%%%%%%%%%%%%%%%%%%%%%%%%%%%%%%%%%%%%%

\par
\section{Supplementary: A Brief Survey of the Topology of Anti-BRST Structure}
\setcounter{equation}{0}

\par In this section we aim to have a short review on the geometric and the topological structures of anti-BRST due to our given model. In order to understand the topology of BRST and anti-BRST symmetries, we initially wish to work out some facts about the topological structures of $\widetilde{\mathcal G}$ and $\widetilde{\widetilde{\mathcal G}}$ via their de Rham cohomology groups. At the first step, it should be noted that in $n$-dimensional spacetime, the gauge transformation group $\widetilde{\mathcal G}$ is decomposed to connected components for each element of the $n$-th homotopy group $\pi_n(\mathcal G)$,\footnote{Actually, this decomposition takes place since the gauge transformations are subjected to one point preserving condition at the infinity point of the compact spacetime $S^n=\mathbb{R}^n \cup \{\infty\}$.} i.e. $\widetilde{\mathcal G}=\overline{\bigcup}_s \widetilde{\mathcal G}_s$, for $s \in \pi_n(\mathcal G)$.\footnote{In this paper the symbol of $\overline{\bigcup}$ is utilized for disjoint unioun.} Hence;
\begin{equation} \label {jadid}
H^*_{\text{de Rham}}(\widetilde{\mathcal G},\mathbb{R}) \cong \oplus_s H^*_{\text{de Rham}}(\widetilde{\mathcal G}_s,\mathbb{R})  \cong \oplus_s H^*_{\text{de Rham}}(\widetilde{\mathcal G}_0,\mathbb{R}) 
\end{equation}
\noindent wherein $\widetilde{\mathcal G}_0$ is the connected component of $\widetilde{\mathcal G}$ including the identity element. It is worth to note that in the second isomorphism we used the homeomorphism $\widetilde{\mathcal G}_s \simeq \widetilde{\mathcal G}_{s'}$,\footnote{We employ the symbol of $\simeq$ for homotopy (topologically) equivalent spaces.} for any $s,s' \in \pi_n(\mathcal G)$.

\par In addition, $\widetilde{\mathcal X}=\widetilde{\mathcal P}/\widetilde{\mathcal G}$ consists of connected components labeled by instanton numbers $k$, i.e. the winding numbers of the homotopic maps $\hat{g}:S^{n-1} \to \mathcal G$, wherein here $S^{n-1}$ is the equator of the compactified spacetime $\mathcal X=\mathbb{R}^n \cup \{\infty\}=S^{n}$. More precisely, the connected components of $\widetilde{\mathcal X}$ are in fact characterized by the elements of the $(n-1)$-th homotopy group $\pi_{n-1}(\mathcal G)$, i.e. $\widetilde{\mathcal X}=\overline{\bigcup}_{k} {\widetilde{\mathcal X}_k}$, for instaton numbers $k \in \pi_{n-1}(\mathcal G)$. On the other hand, any gauge fixing transformation $\widetilde{\widetilde{g}}:\widetilde{\mathcal X} \to \widetilde{\mathcal G}$ could be considered as a smooth map from $\widetilde{\mathcal X}\times \mathcal X$ to $\mathcal G$, i.e. $\widetilde{\widetilde{g}}:\widetilde{\mathcal X}\times \mathcal X \to \mathcal G$ (which we show it with the same notation $\widetilde{\widetilde{g}}$). Thus; $\widetilde{\widetilde{\mathcal G}}=\prod_k \widetilde{\widetilde{\mathcal G}}_k$, wherein $\widetilde{\widetilde{\mathcal G}}_k=\{\widetilde{\widetilde{g}}:\widetilde{\mathcal X}_k \times \mathcal X \to \mathcal G \}$. Hence, if $\widetilde{\mathcal X}_k$s are contractible (e.g. the case of $\mathcal G=U(1)$), then we find the homotopy equivalences; $\widetilde{\widetilde{\mathcal G}}\simeq \prod_k \widetilde{\mathcal G} \simeq \overline{\bigcup}_s (\prod_k \widetilde{\mathcal G}_0)$, and;
\begin{equation} \label {ks}
H^*_{\text{de Rham}}(\widetilde{\widetilde{\mathcal G}},\mathbb{R})\cong \oplus_s H^*_{\text{de Rham}}( \prod_k  \widetilde{\mathcal G}_0,\mathbb{R}).
\end{equation}

\noindent Otherwise, using the homotopy equivalence $\widetilde{\mathcal X}_k \simeq \widetilde{\mathcal X}_{k'}$, for instanton numbers $k, k' \in \pi_{n-1}(\mathcal G)$, one concludes that $\widetilde{\widetilde{\mathcal G}}=\prod_k \widetilde{\widetilde{\mathcal G}}_k \simeq \prod_k \widetilde{\widetilde{\mathcal G}}_0$. Here, $\widetilde{\widetilde{\mathcal G}}_0$ is the gauge transformation group for some specified connected component of $\widetilde{\mathcal X}$, say $\widetilde{\mathcal X}_0$. In this case, we also find a simple formula for the cohomology groups of $\widetilde{\widetilde{\mathcal G}}$;
\begin{equation} \label{cohomology}
H^*_{\emph{\emph{de Rham}}}(\widetilde{\widetilde{\mathcal G}},\mathbb{R}) \cong  H^*_{\emph{\emph{de Rham}}}(\prod_k \widetilde{\widetilde{\mathcal G}}_0,\mathbb{R}).
\end{equation}

\par As it is shown in the following, studying these cohomology groups shed more light on the geometric and topological aspects of the Yang-Mills theories and their BRST/anti-BRST symmetries. To see this with more details we should revisit the Faddeev-Popov Lagrangian from the cohomological viewpoint. This Lagrangian has the following formal structure;
\begin{equation} \label {lagrangian}
\mathcal L_{\text{F-P}}=\mathcal L_{\text{Y-M}}+\alpha \text{Tr}\{\frak F^2 \}- 2 \text{Tr}\{\widetilde{\mathcal C}{\partial_a \frak F}\}\text{Tr}\{t^a \mathcal C\}- 2 \text{Tr}\{ \mathcal V \frak F\}+\alpha^{-1} \text{Tr}\{\mathcal V^2 \},
\end{equation}

\noindent wherein $\alpha$ is a real number, $t^a$ is the generator of the gauge group $\mathcal G$ with $\text{Tr}\{t^at^b\}=-\frac{1}{2}\delta^{ab}$, $\frak F$ is the gauge fixing function and $\partial_a \frak F=\frac{\delta \frak F(\mathcal A.\widetilde{g})}{\delta \widetilde{g}}$ for $\widetilde{g}=e^{\epsilon t^a}$, with $\epsilon \to 0$. Let us rewrite this Lagrangian as $\mathcal L_{\text{F-P}}=\mathcal L_0+\mathcal L_1+\mathcal L_2$ with;
\begin{equation} \label {lagrangian2}
\mathcal L_0=\mathcal L_{\text{Y-M}}+\alpha \text{Tr}\{\frak F^2 \},~~~~~\mathcal L_1=-2 \text{Tr}\{\widetilde{\mathcal C}{\partial_a \frak F}\}\text{Tr}\{t^a \mathcal C\}-2\text{Tr}\{ \mathcal V \frak F\},~~~~~\mathcal L_2=\alpha^{-1} \text{Tr}\{\mathcal V^2 \}.
\end{equation}

\noindent Actually, due to (\ref{2-17}) and (\ref{2-18}) it is seen that each of these three terms is both BRST and anti-BRST closed form in its own right \cite{BRS,T, Curci, Ojima}, i.e.;
\begin{equation} \label {closedness}
~~~~~~~~~~\delta \mathcal L_k=\widetilde{\delta} \mathcal L_k=0,~~~~~k=0,1,2.
\end{equation}

\noindent These equations indicate that each of the above terms belongs to a cohomology group of the BRST and the anti-BRST complex, i.e. the de Rham cohomology groups of $\widetilde{\mathcal G}$ and $\widetilde{\widetilde{\mathcal G}}$. In principle, with thanks to the geometric formulas of (\ref{2-11}) and (\ref{bb'}), and counting the ghost and the ant-ghost fields included in $\mathcal L_k$s we readily see;
\begin{equation} \label {l_k}
~~~~~~~~~~[\mathcal L_k] \in  R\otimes H^k_{\emph{\emph{de Rham}}}(\widetilde{\mathcal G},\mathbb{R}) \otimes H^k_{\emph{\emph{de Rham}}}(\widetilde{\widetilde{\mathcal G}},\mathbb{R}),~~~~~k=0,1,2,
\end{equation}

\noindent wherein $R$ is the ring of the local cohomology of the spacetime $\mathcal X$, and $[\mathcal L_k]$ is the cohomology class of $\mathcal L_k$. Actually, based on the celebrated Kunneth formula
\begin{equation} \label {kunneth-taze}
H^{m}_{\emph{\emph{de Rham}}}(\widetilde{\mathcal G}\times \widetilde{\widetilde{\mathcal G}};\mathbb{R})
\cong  \oplus_{l=0}^m \left(  H^l_{\emph{\emph{de Rham}}}(\widetilde{\mathcal G},\mathbb{R}) \otimes H^{m-l}_{\emph{\emph{de Rham}}}(\widetilde{\widetilde{\mathcal G}},\mathbb{R})\right),
\end{equation}

\noindent the Lagrangian term $\mathcal L_k$, represents a definite de Rham cohomology class in $H^{2k}_{\emph{\emph{de Rham}}}(\widetilde{\mathcal G}\times \widetilde{\widetilde{\mathcal G}};\mathbb{R})$, for each $k=0,1,2$. For instance, since
\begin{equation} \label{R-cohomology}
[\mathcal L_0] \in R \otimes H^{0}_{\text{de Rham}}(\widetilde{\mathcal G}\times \widetilde{\widetilde{\mathcal G}};\mathbb{R}),
\end{equation}

\noindent then $\mathcal L_0$ must be a constant function on both Lie groups $\widetilde{\mathcal G}$ and $\widetilde{\widetilde{\mathcal G}}$, i.e. it does not vary against gauge and gauge fixing transformations, just as we expected before for $\mathcal L_{\text{Y-M}}$. The invariance of $\text{Tr}\{\frak F^2\}$ with respect to the gauge and gauge fixing transformations, as reported in (\ref {closedness}), is in fact a conclusion of the supersymmetric structure of $\frak F$ as a differential form \cite{Weinberg}.\footnote{Actually, we have;
\begin{align*}
\begin{array}{*{20}{c}}
\delta \text{Tr}\{\frak F^2\}=\text{Tr}\{\frak F'(\text{d}\mathcal C+\mathcal C\mathcal A+\mathcal A\mathcal C)\frak F\}+(-1)^{|\frak F|}\text{Tr}\{\frak F \frak F' (\text{d}\mathcal C+\mathcal C \mathcal A+\mathcal A\mathcal C)\}\\
~~~~~~~~~~~~~~~~~~~~~=\text{Tr}\{\frak F'(\text{d}\mathcal C+\mathcal C\mathcal A+\mathcal A\mathcal C)\frak F\}+(-1)^{2|\frak F|-1}\text{Tr}\{\frak F' (\text{d}\mathcal C+\mathcal C \mathcal A+\mathcal A\mathcal C)\frak F\}=0,
\end{array}
\end{align*}
\noindent wherein $\frak F'=\frac{\delta \frak F}{\delta \mathcal A}$ and $|\frak F|$ is the degree of $\frak F$ as a differential form. The same result holds for the anti-BRST derivative $\widetilde{\delta}$ via replacing $\mathcal C$ with $\widetilde{\mathcal C}$ in the calculation. See \cite{Ojima} for more details.}

\par However, the most significant topological features of the Faddeev-Popov theory (Lagrangian) are essentially encoded in $\mathcal L_1$ and $\mathcal L_2$. But to see the topological significance of these terms we need more precisions about the topology of principal bundle $(\widetilde{\mathcal P},\widetilde{\pi},\widetilde{\mathcal X})$. Based on definitions (\ref {2-7}) to (\ref {2-8}), and (\ref {2-11}), the quantum fields $\mathcal C$, $\widetilde{\mathcal C}$ and $\mathcal V$ are smooth differential forms on $\widetilde{\mathcal G} \times \widetilde{\widetilde{\mathcal G}}$ if and only if the principal bunlde is topologically trivial, provided that it could admit a globally defined smooth section $\widetilde{\frak s}$. Otherwise, these fields are smooth only on a dense open set in $\widetilde{\mathcal G} \times \widetilde{\widetilde{\mathcal G}}$. More precisely, for a non-trivial topology of $\widetilde{\pi}:\widetilde{\mathcal P} \to \widetilde{\mathcal X}$, the geometric elaborations we worked out in the previous section lead to almost everywhere smooth structures, i.e. the formulations which are smooth outside a null subset of the underlying manifold.

\par Actually, such a precision was not important so far from the algebraic perspective to the BRST-anti-BRST symmetry. Even, one can ignore this accuracy for computing the integral of a functional of such quantum fields on the underlying manifold (such as the Faddeev-Popov action which is given by an integral on the spacetime\footnote{In principle, here we may also ignore the non-trivial topology of $(\mathcal P,\pi,\mathcal)$ for compact spacetime $\mathcal X=S^n$.}), since in fact, the contribution of null sets in integration is always nil. But, in contrast to the calculus and the algebraic formulations, this precision plays the most substantial role for studying the topology of the anti-BRST structure.

\par Now, by considering the non-triviality of the topology of $(\widetilde{\mathcal P},\widetilde{\pi},\widetilde{\mathcal X})$ we can introduce two topological invariants of the quantized theory based on the Lagrangian terms $\mathcal L_1$ and $\mathcal L_2$.  As we mentioned above for the former we have;
\begin{equation} \label{faddeev-popov}
[\mathcal L_1] \in R\otimes H^1_{\emph{\emph{de Rham}}}(\widetilde{\mathcal G},\mathbb{R}) \otimes H^1_{\emph{\emph{de Rham}}}(\widetilde{\widetilde{\mathcal G}},\mathbb{R})\subset R\otimes H^2_{\emph{\emph{de Rham}}}(\widetilde{\mathcal G}\times \widetilde{\widetilde{\mathcal G}},\mathbb{R}).
\end{equation}

\noindent Hence, for any Riemann surface $\Sigma$ and any smooth map $\tau:\Sigma \to \widetilde{\mathcal G}\times \widetilde{\widetilde{\mathcal G}}$ the integral of
\begin{equation} \label {invariant}
\sigma_1=\int_{\Sigma} \tau^*(\mathcal L_1) \in R
\end{equation}

\noindent provides a topological invariant of the theory, i.e. the local cohomology class of the integral is preserved against any continuous variation of $\tau$. We refer to $\sigma_1$ as the \emph{first Nakanishi-Lautrup invariant} of the theory. Due to the structure of $\mathcal L_1$ in (\ref{lagrangian2}), it is obvious that $\sigma_1$ encodes some topological features of the gauge fixing mechanism imposed by $\frak F$ and the Nakanishi-Lautrup field $\mathcal V$, as a local cohomology class of $R$.

\par It is well known that $\mathcal L_1$ is itself a BRST derivation; $\mathcal L_1=\delta \text{Tr}\{2\widetilde{\mathcal C} \frak F\}$ \cite{BRS, T}. Hence, if $\frak F$ is a globally smooth function provided that $(\widetilde{\mathcal P},\widetilde{\pi},\widetilde{\mathcal X})$ is topologically trivial, then $\mathcal L_1$ would be an exact term and $\sigma_1=0$. But, if $(\widetilde{\mathcal P} , \widetilde{\pi} \widetilde{\mathcal X})$ is non-trivial, then $\frak F$ would not be a globally smooth function (as the Gribov ambiguity confirms), and consequently the cohomology class of $\mathcal L_1$ in $H^1_{\text{de Rham}} (\widetilde{\mathcal G},\mathbb{R})$ would not be null. On the other hand, it can be seen that $\mathcal L_1=\delta \widetilde{\delta} \widetilde{\frak F}$, wherein $\widetilde{\frak F}$ is a function from $\widetilde{\mathcal P}\times \widetilde{\widetilde{\mathcal G}}$ to $\mathbb{R}$ with the property of $\widetilde{\delta} \widetilde{\mathcal F}=\text{Tr}\{2\widetilde{\mathcal C}\frak F\}$ \cite{Curci, Ojima}. Therefore, if $\frak F$ is not globally smooth, neither is $\widetilde{\frak F}$, and in conclusion $\mathcal L_1$ defines a non-trivial cohomology class of $H^2_{\text{de Rham}}(\widetilde{\mathcal G}\times \widetilde{\widetilde{\mathcal G}},\mathbb{R})$.\footnote{We should note that this is a standard method for defining non-trivial cohomology classes on smooth manifolds. For instance, $\text{d}_1\text{d}_2 (\theta_1 \theta_2)=\text{d}_1\theta_1 \wedge \text{d}_2 \theta_2$ defines a non-trivial cohomology class of $H^2_{\text{de Rham}}(S^1\times S^1,\mathbb{R})$, whereas $\theta_1 \theta_2$ is a discontinuous function on $S^!\times S^1$. Here, $\theta_i$ is the canonical angle of the $i$-th copy of $S^1$ with the exterior derivative $\text{d}_i$, $i=1,2$.} Thus, upon the Gribov ambiguity, the first Nakanishi-Lautrup invariant provides a non-trivial topological information of the quantized Yang-Mills theory, i.e. generally we obtain $\sigma_1\ne 0$.

\par On the other hand, we also have;
\begin{equation} \label{faddeev-popov}
[\mathcal L_2] \in R\otimes H^2_{\emph{\emph{de Rham}}}(\widetilde{\mathcal G},\mathbb{R}) \otimes H^2_{\emph{\emph{de Rham}}}(\widetilde{\widetilde{\mathcal G}},\mathbb{R})\subset R\otimes H^4_{\emph{\emph{de Rham}}}(\widetilde{\mathcal G}\times \widetilde{\widetilde{\mathcal G}},\mathbb{R}).
\end{equation}

\noindent Hence, for any closed 4-manifold $M$ and any smooth map $\tau:M \to \widetilde{\mathcal G}\times \widetilde{\widetilde{\mathcal G}}$ the integral of
\begin{equation} \label {invariant-2}
\sigma_2=\int_{M} \tau^*(\mathcal L_2) \in R
\end{equation}

\noindent produces another topological invariant of the theory, i.e. the local cohomology class of the integral is preserved versus any continuous variation of $\tau$. We also refer to $\sigma_2$ as the \emph{second Nakanishi-Lautrup invariant} of the theory. But, actually we can see that $\mathcal L_2=\delta \text{Tr}\{\widetilde{\mathcal C} \mathcal V\}=\widetilde{\delta} \text{Tr}\{\mathcal C \mathcal V \}$ \cite{BRS, T, Ojima}. Therefore, as we explained above, if the topology of $(\widetilde{\mathcal P},\widetilde{\pi},\widetilde{\mathcal X})$ is trivial, then $[\mathcal L_2]=0 \in H^4_{\emph{\emph{de Rham}}}(\widetilde{\mathcal G}\times \widetilde{\widetilde{\mathcal G}},\mathbb{R})$ and hence; $\sigma_2=0$. Otherwise, for non-trivial topology of the principal bundle, once again we obtain $\sigma_2\ne 0$.

\par It is worth to emphasis that these two new topological information of the theory, the Nakanishi-Lautrup invariants $\sigma_1$ and $\sigma_2$, owe their existence to the geometric structures we worked out in the previous section. In fact, they contain absolutely new topological information from the theory based on the gauge fixing symmetry structure, which had so far been hiding behind the mysterious appearance of the Nakanishi-Lautrup field.

\par One can also employ the anti-ghost consistent anomalies to work out more topological invariants of the theory based on the topology of $\widetilde{\widetilde{\mathcal G}}$ cite{Varshovi1}. It is in fact an element of $H^1_{\emph{\emph{de Rham}}}(\widetilde{\widetilde{\mathcal G}},\mathbb{R})$ and can be used to study the topological aspects of the anti-BRST structure. Actually, the procedure we introduced in \cite {Varshovi1} to write down the BRST/anti-BRST descent equations (and to work out the ghost/anti-ghost consistent anomalies) holds similarly for the generalized BRST-anti-BRST algebra we worked out here via (\ref{2-17}) and (\ref{2-18}). Thus, we obtain the same formulas for both the ghost consistent anomaly $\frak G_{\text{BRST}}(\mathcal A)$, and the anti-ghost consistent anomaly $\widetilde{\frak G}_{\text{anti-BRST}}(\mathcal A)$ in four dimensional spacetime, i.e.;\footnote{In fact, proving equations of (\ref{consistent}) is based on the three first equalities in (\ref{2-17}) and (\ref{2-18}). Therefore, the details of the proof are exactly as discussed in \cite{Varshovi1}.}
\begin{equation} \label {consistent}
\begin{array} {*{20}{c}}
\frak G_{\text{BRST}}(\mathcal A)=\kappa ~\text{Tr}\{\text{d}\mathcal C (\mathcal A \text{d}\mathcal A +\frac{{1}}{{2}} \mathcal A^3)\},\\
\widetilde{\frak G}_{\text{anti-BRST}}(\mathcal A)=\kappa ~\text{Tr}\{\text{d}\widetilde{\mathcal C} (\mathcal A \text{d}\mathcal A +\frac{{1}}{{2}} \mathcal A^3)\},~~~~
\end{array}
\end{equation}

\noindent with $\kappa=\frac{1}{24\pi^2}$. These two differential forms define specific cohomology classes of $R \otimes H^1_{\emph{\emph{de Rham}}}({\widetilde{\mathcal G}},\mathbb{R})$ and of $R \otimes H^1_{\emph{\emph{de Rham}}}(\widetilde{\widetilde{\mathcal G}},\mathbb{R})$ respectively. Actually, $\widetilde{\frak G}_{\text{anti-BRST}}(\mathcal A)$ itself defines a topological invariant of the theory based on the topology of $\widetilde{\widetilde{\mathcal G}}$. In fact, for any smooth $\widetilde{\tau}:S^1 \to \widetilde{\widetilde{\mathcal G}}$, the outcome of
\begin{equation} \label{index-0}
\sigma_3=\int_{S^1\times {\mathcal X}} {\widetilde{\tau}}^* (\widetilde{\frak G}_{\text{anti-BRST}}(\mathcal A)) \in \mathbb{Z}
\end{equation}

\noindent is also a topological invariant, which is preserved via any continuous variation of $\widetilde{\tau}$ and the action of the gauge fixing transformation group $\widetilde{\widetilde{\mathcal G}}$. Accordingly, we refer to $\sigma_3$ as the \emph{anti-BRST topological index}.

\par Similar structures of $\frak S_{\text{BRST}}(\mathcal A)$ and $\widetilde{\frak S}_{\text{anti-BRST}}(\mathcal A)$ reveals that the BRST symmetry and the anti-BRST invariance gain similar topological aspects for general gauge fixing methods. In principle, the similarity of the BRST and the anti-BRST structures derived for the strict gauges in \cite{Varshovi1}, also hold for the smooth gauge fixing mechanisms. For instance, we find the same topological index for the BRST and the anti-BRST symmetries;
 \begin{equation} \label{index}
\int_{S^1\times {\mathcal X}} \tau^*(\frak G_{\text{BRST}}(\mathcal A))=\int_{S^1\times {\mathcal X}} {\widetilde{\tau}}^* (\widetilde{\frak G}_{\text{anti-BRST}}(\mathcal A)) \in \mathbb{Z},
\end{equation}
 \noindent wherein here $\widetilde{\tau}:S^1 \to \widetilde{\widetilde{\mathcal G}}$ is the composition of $\tau$ and the natural inclusion $\widetilde{\mathcal G} \subset \widetilde{\widetilde{\mathcal G}}$.\footnote{See \cite{Varshovi1} for more details.}
 
 \par Also for the case of smooth gauge fixing mechanisms the BRST and the anti-BRST symmetries give rise to two Noether's currents with corresponding charges $\frak S$ and $\widetilde{\frak S}$. With thanks to (\ref {13}), which hold generally, these are two distinct nilpotent supersymmetric operators on the Hilbert space of the gauge field quantum states. Therefore, for any gauge fixing mechanism, they provide two cohomology groups defined as $H_{\text{BRST}}(\frak S):=\text{Im}(\frak S)/\text{Ker}(\frak S)$ and $H_{\text{anti-BRST}}( \widetilde{\frak S}):=\text{Im}(\widetilde{\frak S})/\text{Ker}(\widetilde{\frak S})$. We can prove that these groups are still comparable in the presence of the Nakanishi-Lautrup field. All in all, we have the following theorem; \\

\par \textbf{Theorem 3 (Generalized Version of \cite{Varshovi1})\textbf{;}}\footnote{All the statements of \textbf{Theorem 3} are true for faith-full irreducible representations of gauge groups and specially for fundamental representations \cite{Bonora et al}.} \emph{In a quantized Yang-Mills theory with generally non-vanishing Nakanishi-Lautrup fields (with either sharp or smooth gauge fixing mechanisms) the BRST symmetry is broken if and only if the anti-BRST invariance is lost. More precisely, after quantizing a Yang-Mills theory the gauge fixing invariance would be anomalous if and only if the gauge symmetry is anomalous too. On the other hand, the cohomology groups of the BRST and the anti-BRST charges coincide for any arbitrary gauge fixing mechanism, i.e.;}
\begin{equation} \label {brst-cohomology}
H_{\text{BRST}}(\frak S) \cong H_{\text{anti-BRST}}(\widetilde{\frak S}).\\
\end{equation}
~\\
 \indent \textbf{Proof;} The first assertion of the theorem is an immediate consequence of (\ref{consistent}) or (\ref{index}). The equality of the BRST and the anti-BRST cohomology groups is also the result of the Faddeev-Popov Lagrangian. Actually, since the Nakanishi-Lautrup field has no dynamics it gives no contribution to the Hilbert space of the theory. Thus, it is seen that for the smooth gauge fixing mechanisms the cohomology groups $H_{\text{BRST}}(\frak S)$ and $H_{\text{anti-BRST}}(\widetilde{\frak S})$ coincide with their counterparts derived for the strict gauges. But, on the other hand, it was proved in \cite{Varshovi1} that these groups are equal for the strict gauges. This finishes the theorem. \textbf{Q.E.D.}

%%%%%%%%%%%%%%%%%%%%%%%%%%%%%%%%%%%%%%%%%%%%%%%%%%%%%
%%%%%%%%%%%%%%%%%%%%%%%%%%%%%%%%%%%%%%%%

\par
\section{Summary and Conclusion}
\setcounter{equation}{0}
\par Through this research we worked out a geometric formulation for anti-BRST structure of quantized Yang-Mills gauge theories based on the symmetry of the classical theories with respect to gauge fixing methods. Here we established that this symmetry, for any general gauge fixing mechanism, including either the strict gauges (such as that of Landau) or the smooth ones (such as that of Feynman), appears as the classical counterpart of anti-BRST invariance. Hence, it was also demonstrated that for any quantized Yang-Mills theory the anti-BRST symmetry is an obligation just as the BRST invariance. That is, emerging the anti-BRST symmetry via the Faddeev-Popov quantization approach is as natural as appearing the gauge fixing invariance of the classical theory. Thus, this geometric model explains the mysterious appearance of the anti-BRST symmetry beside that of the BRST in the Faddeev-Popov Lagrangian.\footnote{As was noted by Alvarez-Gaume \cite{Alvarez-Gaume} who believed that the anti-BRST symmetry has no classical counterpart.}

\par In addition to the geometric structures of the gauge fixing symmetry, we studied its topological aspects in the setting of the de Rham cohomology of the gauge fixing transformation group. We showed that the gauge fixing terms appearing in the Faddeev-Popov Lagrangian are classified via the topological effects of the instantonic structures of the Yang-Mills theory. Then three topological invariants due to the gauge fixing symmetry were introduced for Yang-Mills theories, the \emph{first Nakanishi-Lautrup invariant} $\sigma_1$, the \emph{second Nakanishi-Lautrup invariant} $\sigma_2$, and the \emph{anti-BRST topological index} $\sigma_3$. Nest, we argued that the BRST and the anti-BRST anomalous structures, the ghost and the anti-ghost consistent anomalies, coincide generally for any gauge fixing mechanism. More precisely, we proved that no matter which gauge fixing approach is employed to quantize a Yang-Mills theory, the BRST symmetry (gauge symmetry) is broken at quantum levels if and only if the anti-BRST invariance (gauge fixing invariance) is lost too. Finally, we established that the BRST and the anti-BRST cohomology groups will coincide for general gauges.
\\

%%%%%%%%%%%%%%%%%%%%%%%%%%%%%%%%%%%%%%%%%%%%%%%%
\section{Acknowledgments}
%%%%%%%%%%%%%%%%%%%%%%%%%%%%%%%%%%%%%%%%%%%%%%%%
\par The author says his special gratitude to S. Ziaee who was the main reason for appearing this article. Moreover, it should be noted that  this research was in part supported by a grant from IPM (No.99810422).

%%%%%%%%%%%%%%%%%%%%%%%%%%%%%%%%%%%%%%%%%%%%%%%%%%%%%%%%%%%%%%%%%%
%\section{Appendices}
%%%%%%%%%%%%%%%%%%%%%%%%%%%%%%%%%%%%%%%%%%%%%%%%%%%%%%%%%%%%%%%%%%%
%\begin{appendix}\setcounter{equation}{0}\noindent
%\section{}\

%%%%%%%%%%%%%%%%%%%%%%%%%%%%%%%%%%%%%%%%%

\end{document}